\documentclass[a4paper,11pt]{article}
\usepackage{jheppub} 
\pdfoutput=1
\usepackage{graphicx}
\usepackage{makecell,booktabs,array}
\newcolumntype{x}[1]{%
>{\centering\hspace{0pt}}p{#1}}%

\usepackage{amsmath}
\usepackage{amssymb}
\usepackage{subcaption}

\newcommand{\dmse}{\ensuremath{d^{(2)}_{\rm{MPE}}}}
\newcommand{\demd}{\ensuremath{d^{(1)}_{\rm{Wass}}}}

\usepackage{graphicx}% Include figure files
\usepackage{dcolumn}% Align table columns on decimal point
\usepackage{bm}% bold math
\usepackage{multirow}

\usepackage{color,soul}
\usepackage{tikz}
\usepackage{hyperref}

\definecolor{darkyellow}{rgb}{0.5, 0.5, 0.0}
\definecolor{darkpurple}{rgb}{0.5, 0.2, 0.8}
\definecolor{darkblue}{rgb}{0.0, 0.0, 0.8}
\definecolor{darkgreen}{rgb}{0.0, 0.4, 0.0}
\definecolor{darkred}{rgb}{0.5, 0.0, 0.0}
\usepackage{hyperref}
\hypersetup{
    linktocpage,
     colorlinks,
     citecolor=darkblue,
     linkcolor= darkblue,
     urlcolor=darkblue
}

\newcommand{\num}{{}}

\title{Challenges for Unsupervised Anomaly Detection in Particle Physics}

\author{Katherine Fraser,}
\author{Samuel Homiller,}
\author{Rashmish K. Mishra,}
\author{Bryan Ostdiek, and }
\author{Matthew D. Schwartz}

\affiliation{Department of Physics, Harvard University, Cambridge, MA 02138, USA}
\affiliation{The NSF AI Institute for Artificial Intelligence and Fundamental Interactions}

\emailAdd{kfraser@g.harvard.edu}
\emailAdd{shomiller@g.harvard.edu}
\emailAdd{rashmishmishra@fas.harvard.edu}
\emailAdd{bostdiek@g.harvard.edu}
\emailAdd{schwartz@g.harvard.edu}

\abstract{
Anomaly detection relies on designing a score to determine whether a particular event is uncharacteristic of a given background distribution.
One way to define a score is to use autoencoders, which rely on the ability to reconstruct certain types of data (background) but not others (signals).
In this paper, we study some challenges associated with variational autoencoders, such as the dependence on hyperparameters and the metric used, in the context of anomalous signal (top and $W$) jets in a QCD background.
We find that the hyperparameter choices strongly affect the network performance and that the optimal parameters for one signal are non-optimal for another. In exploring the networks, we uncover a connection between the latent space of a variational autoencoder trained using mean-squared-error and the optimal transport distances within the dataset. 
We then show that optimal transport distances to representative events in the background dataset can be used directly for anomaly detection, with performance comparable to the autoencoders. 
Whether using autoencoders or optimal transport distances for anomaly detection, we find that the choices that best represent the background are not necessarily best for signal identification. These challenges with unsupervised anomaly detection bolster the case for additional exploration of semi-supervised or alternative approaches.
}

\begin{document}
\maketitle
\flushbottom

%%%%%%%%%%%%%%%%%%%%%%%%%%%%%%%%%%%%%%%%%%%%%%%%%%%%%%%%
\section{Introduction}
\label{sec:intro}
%%%%%%%%%%%%%%%%%%%%%%%%%%%%%%%%%%%%%%%%%%%%%%%%%%%%%%%%

While many searches for physics beyond the Standard Model have been carried out at the Large Hadron Collider, new physics remains elusive. This may be due to a lack of new physics in the data, but it could also be due to us looking in the wrong place. Trying to design searches that are more robust to unexpected new physics has inspired a lot of work on anomaly detection using unsupervised methods including community wide challenges such as the LHC Olympics~\cite{Kasieczka:2021xcg} and the Dark Machines Anomaly Score Challenge~\cite{Aarrestad:2021oeb}. The goal of anomaly detection is to search for events which are ``different'' than what is expected. 
When used for anomaly detection, unsupervised methods attempt to characterize the space of background events in some way, independent of signal. The hope is then that signal events will stand out as being uncharacteristic.

Anomaly detection techniques can be broadly split into two categories. For some signals, the signal events look similar to background events and one must exploit information about the expected probability distribution of the background to find the signal. Many anomaly detection techniques have been developed to find signals of this type~\cite{Collins:2018epr,DAgnolo:2018cun,DeSimone:2018efk,2018arXiv180902977C,Collins:2019jip,Dillon:2019cqt,Mullin:2019mmh,DAgnolo:2019vbw,Nachman:2020lpy,Andreassen:2020nkr,ATLAS:2020iwa,Dillon:2020quc,Benkendorfer:2020gek,Mikuni:2020qds,Stein:2020rou,Kasieczka:2021xcg,Batson:2021agz,Blance:2021gcs,Bortolato:2021zic,Collins:2021nxn,Dorigo:2021iyy,Volkovich:2021txe,Hallin:2021wme}.
Alternatively, some signals are qualitatively different from background and then methods that try to characterize an individual event as anomalous can be used~\cite{Aguilar-Saavedra:2017rzt,Hajer:2018kqm,Heimel:2018mkt,Farina:2018fyg,Cerri:2018anq,Roy:2019jae,Blance:2019ibf,RomaoCrispim:2019tai,Amram:2020ykb,CrispimRomao:2020ejk,Knapp:2020dde,CrispimRomao:2020ucc,Cheng:2020dal,Khosa:2020qrz,Thaprasop:2020mzp,Aguilar-Saavedra:2020uhm,Pol:2020weg,vanBeekveld:2020txa,Park:2020pak,Chakravarti:2021svb,Faroughy:2020gas,Finke:2021sdf,Atkinson:2021nlt,Dillon:2021nxw,Kahn:2021drv,Aarrestad:2021oeb,Caron:2021wmq,Govorkova:2021hqu,Gonski:2021jek,Govorkova:2021utb,Ostdiek:2021bem}. Here, we restrict to the latter type of anomaly detection, where an anomaly
score for individual events can be used for discrimination, without needing to characterize the full probability distribution of the signal ensemble. With an effective method, events with a small score are likely to be a part of the background distribution, while events with a larger score are not. There are many different ways of defining an anomaly score. Some rely on traditional high-level observables, like mass or $N$-subjettiness. Others attempt to directly learn how likely a given event or object is using low-level information, like individual particle momenta (see e.g., ref.~\cite{Caron:2021wmq}).
Some methods that search for outliers rely on abstract representations to try to characterize the event space, such as the latent space of an autoencoder \cite{Dillon:2021nxw, Bortolato:2021zic}. Others give the event space itself a geometric interpretation in terms of distances~\cite{Komiske:2019fks, Komiske:2020qhg, Cai:2020vzx}. Given the complexity and high-dimensionality of data at the LHC, many anomaly detection techniques employ machine learning.

In this paper, we begin by exploring the use of autoencoders for anomaly detection.
Autoencoders were initially introduced for dimensionality reduction, similar to principal component analysis, to learn the important information in data while ignoring insignificant information and noise~\cite{https://doi.org/10.1002/aic.690370209}. Autoencoders contain an encoder, which reduces the dimensionality of the input to give some latent representation, and a decoder, which transforms the latent space back to the original space.
In particle physics, autoencoders were first used for anomaly detection in refs.~\cite{Farina:2018fyg, Heimel:2018mkt}, where they are meant to reconstruct certain types of data (background) but not others (signals).  
In order to work as an anomaly detector, an autoencoder should have a small reconstruction error for background events and a large reconstruction error for signal events. To do so, the autoencoder must establish a delicate balance in achieving a reconstruction fidelity which is accurate, but not too accurate. There are several cases where this is especially difficult, such as when the signal-to-background ratio $S/B$ is small, when the dataset has certain topological properties~\cite{Batson:2021agz}, or when innate characteristics of the samples make the signal sample simpler than the background sample to reconstruct~\cite{Dillon:2021nxw,Finke:2021sdf}.

A generalization of autoencoders called variational autoencoders (VAEs) were introduced in ref.~\cite{2013arXiv1312.6114K}. Unlike an ordinary autoencoder, where each input is mapped to an arbitrary point in the latent space, in a VAE, the latent space is a probability distribution which is sampled and mapped back to the original space by the decoder. In addition to the usual reconstruction error, the VAE loss also includes a Kullback-Leibler (KL) divergence component that pushes the latent space towards a Gaussian prior and regularizes network training.
The latent space of the VAE encodes the probability distribution of the background training sample, which can be used in the anomaly score.
VAEs were first used in anomaly detection in computer science in ref.~\cite{An2015VariationalAB}, and first used for particle physics anomaly detection in ref.~\cite{Cerri:2018anq}. They have been widely studied since then \cite{Carrazza:2019cnt, Cheng:2020dal,Dohi:2020eda, Pol:2020weg, vanBeekveld:2020txa, Park:2020pak, Bortolato:2021zic, Dillon:2021nxw, Govorkova:2021utb, Ostdiek:2021bem}.

The task of an autoencoder, variational or not, for unsupervised anomaly detection is to provide a strong universal signal/background discriminant for a variety of signals having access only to background for training.
In principle, this approach is advantageous because it opens the possibility to bypass Monte Carlo simulations and work directly with experimental data, which is almost completely background.
The autoencoder paradigm is based on the vision that there is trade-off between efficacy and generality: the ideal discriminant for a given signal and given background would be ineffective for a different signal and different background while a general discriminant, like the autoencoder, would work decently on a broad class of signals and backgrounds.
The ideal assumes, first, that such a general discriminant exists with an appropriate use case, and second that it can be found by training purely on one or more background samples without any direct information about the signal. However, one has reason to be suspicious: machine learning methods work great at optimizing a given loss, which is meant to correlate strongly with the problem one is trying to solve. For autoencoder anomaly detection, the optimization (background only) is not aligned with the ultimate problem of interest (signal discovery over background), so it should not be surprising if the autoencoder does poorly. In section~\ref{sec:vae-results}, we explore the challenges induced by trying to optimize a VAE in a model agnostic way.

In order to understand what a VAE is learning, we study its latent space. In particular, we look at distance between events in VAE latent space (see~\cite{Dillon:2021nxw, Collins:2021pld} for other studies of VAE latent spaces in particle physics).
Since we can think of the VAE anomaly score as a ``distance” encoding how far any given event is from the background distribution, it is also natural to ask about the distances between individual events. We find there is a significant correlation between the Euclidean distance between events represented in the VAE latent space and the Wasserstein optimal transport distance between events represented as images. We study Wasserstein distances in particular because they were physically motivated in refs. ~\cite{Komiske:2019fks, Komiske:2020qhg, Cai:2020vzx}.

The correlation we observe between distances in the VAE latent space and between the event images motivates us to explore using optimal transport distances between events to define an anomaly score in section~\ref{sec:metrics-based-ad}. One method for using distances directly is to identify representative events in the background sample, and use an event-to-event distance between a given event and the representative event as the score. 
The advantages of this method we propose are that it does not require training a neural network and that it is easily adaptable to different background samples.

This paper is organized as follows. In section~\ref{sec:data}, we provide  information about the dataset used in our study. In section~\ref{sec:definitions}, we provide relevant background information on the metrics used (section~\ref{sec:space-of-metrics}), and the details of the VAE architecture (section~\ref{sec:network}). In section~\ref{sec:vae-results}, we explore the effectiveness of an image based convolutional autoencoder for anomaly detection, including its sensitivity to hyperparameters. We also explore correlations between Euclidean distances in the autoencoder's latent space and optimal-transport distances among the event images in section~\ref{sec:metrics-and-ls}. This motivates the development of methods that directly use the optimal transport distances among events as an alternative to VAEs in section~\ref{sec:metrics-based-ad}. We conclude in section~\ref{sec:conc}.

%%%%%%%%%%%%%%%%%%%%%%%%%%%%%%%%%%%%%%%%%%%%%%%%%%%%%%%%
\section{Datasets}
\label{sec:data}
%%%%%%%%%%%%%%%%%%%%%%%%%%%%%%%%%%%%%%%%%%%%%%%%%%%%%%%%

We begin by describing the datasets we use for our analysis. For concreteness, we focus on anomaly detection in simulated jet events at the LHC.
We will consider QCD dijet events as the background, and consider both top and $W$ jets as representatives of anomalous signal events.
The authors of ref.~\cite{Cheng:2020dal} have provided a suite of jets for Standard Model and Beyond the Standard Model particle resonances which are available on Zenodo~\cite{cheng_taoli_2021_4614656}.
A sample of QCD dijet background events are also provided on Zenodo using the same selection criteria, showering, and detector simulation parameters~\cite{leissner_martin_julien_2020_4641460}.
The datasets were generated with \textsc{MadGraph}~\cite{Alwall:2014hca} and \textsc{Pythia8}~\cite{Sjostrand:2014zea} and used \textsc{Delphes}~\cite{deFavereau:2013fsa} for fast detector simulation.
Jets were clustered using \textsc{FastJet}~\cite{Cacciari:2011ma,Cacciari:2005hq} using the anti-$k_T$ algorithm~\cite{Cacciari:2008gp} with a cone size of $R=1.0$.
The event selection requires two hard jets, with leading jet having $p_T > 450$ GeV and the sub-leading jet having $p_T > 200$ GeV.
The QCD jets are created using the $pp\rightarrow jj$ process in \textsc{MadGraph}, while the top and $W$ jets we examine are produced through a $Z^{\prime}$ which decays to $t\bar{t}$ or a $W^{\prime}$ which decays to $W + jj$, respectively.
There are around 700,000 QCD dijet events and 100,000 events for the ``anomalous'' top and $W$ events.
We reserve 100,000 QCD events for testing and use 50,000 QCD events for validation when training the VAE.

The leading jet in each event is used for the analysis.
We pre-process the raw four-vectors into an image following the procedure presented in ref.~\cite{Macaluso:2018tck}.
Using the \textsc{EnergyFlow} package~\cite{energyflow}, we boost and rotate the jet along the beam direction so that the $p_T$ weighted centroid is located at $(\eta, \phi) = (0,0)$.
Next, the jet is rotated about the centroid such that the $p_T$ weighted major principal axis is vertical.
After this, the jet is flipped along both the horizontal and vertical axes so that the maximum intensity is in the upper right quadrant.
Only after the centering, rotations, and flipping do we pixelate the data~\cite{Macaluso:2018tck}.
We use $40\times40$ pixel images covering a range of $\Delta \eta = \Delta \phi = 3.2$.
The final step of the pre-processing is to divide by the total $p_T$ in the image.
Note that we do \emph{not} standardize each pixel by, e.g., subtracting the mean and dividing by the standard deviation for the entire training dataset, because optimal transport requires positive values in every pixel. It is important to note that the individual images are very sparse and do not resemble the average of the dataset. For instance, out of the 1600 pixels, only $10.4 \pm 5.3$, $13.5 \pm 4.3$, and $10.1 \pm 3.3$ pixels account for more than $1\%$ of the total $p_T$ of the image for the QCD, top, and $W$ jets, respectively.

%%%%%%%%%%%%%%%%%%%%%%%%%%%%%%%%%%%%%%%%%%%%%%%%%%%%%%%%
\section{Defining the Anomaly Score}
\label{sec:definitions}
%%%%%%%%%%%%%%%%%%%%%%%%%%%%%%%%%%%%%%%%%%%%%%%%%%%%%%%%

Anomaly detection, in general, requires an anomaly score: we want to determine if an event is anomalous by measuring how far away it is from the elements of the background distribution. This anomaly score can also be thought of as the ``distance'' between an event and an ensemble. In order to define an event-to-ensemble distance it is helpful first to explore event-to-event distance measures. For instance, given an event-to-event metric, one could compute the distance from an event to some fiducial background event, and use this as a proxy for the event-to-ensemble distance. To understand both types of distances, we need to review the metrics used to define the distance, which we will do in Section~\ref{sec:space-of-metrics}. We can also use an autoencoder to generate an implicit construction of an approximate event-to-ensemble distance, in the form of an anomaly score. We will provide background and discuss the architecture of our autoencoder in section~\ref{sec:network}.

%%%%%%%%%%%%%%%%%%%%%%%%%%%%%%%%%%%%%%%%%%%%%%%%%%%%%%%%
\subsection{Metrics}
\label{sec:space-of-metrics}
%%%%%%%%%%%%%%%%%%%%%%%%%%%%%%%%%%%%%%%%%%%%%%%%%%%%%%%%

First, we define the metrics that can be used to compute event-to-event distances. One of the simplest event representations is to treat an event as an image, with pixel intensities representing the particles' energies~\cite{Cogan:2014oua}.\footnote{
In principle, it would be interesting to consider the complete set of four-vectors of the particles in an event as a representation, rather than the pixelated image, and define a metric on these. 
The $p$-Wasserstein distances described later in this section are well-suited for such a representation, but
building an autoencoder architecture on the full set of four-vectors is more challenging.
It is also important to comment that our image representation is dependent on its preprocessing. Although recent studies have shown that the processing done to events before anomaly detection is inherently model dependent~\cite{Finke:2021sdf}, we work with the images as described.
}
A simple event-to-event metric, the ``mean power error'' (MPE), can then be written as:
\begin{equation}
    d^{(\alpha)}_{\rm{MPE}}(\mathcal{I}_1, \mathcal{I}_2) = \frac{1}{N_{\rm{pixels}}}\sum_{i\in\rm{pixels}} \left|\mathcal{I}_{1,i} - \mathcal{I}_{2,i} \right|^\alpha\:.
    \label{eq:MSE-generalization}
\end{equation}
where $\mathcal{I}_{1(2), i}$ is the pixel intensity (transverse momentum) in pixel $i$ of the image 1(2), and $\alpha$ is a parameter that governs the relative importance of pixels with high/low intensity differences.
This type of metric is often used for doing regression. Frequently, the choice $\alpha = 2$ is made, inspired by the $\chi^2$ statistic, in which case $d^{(2)}_{\rm{MPE}}$ is known as the mean-square error (MSE). The mean-absolute error (MAE) is another well-known choice, corresponding to $\alpha = 1$.

While $d^{(\alpha)}_{\rm{MPE}}$ makes sense in regression, using it on images does not make much sense from a physics point of view.\footnote{In contrast, if one designs a neural network with higher-level variables as the input data representation, using MPE as the metric is a sensible choice.}
For instance, let $\mathcal{I}_1$ be the image of a particle with energy $E$ in a single pixel and $\mathcal{I}_2$ be the image of a particle with same energy $E$ in the neighboring pixel. These events are nearly identical physically, but will have a  very large MSE distance. Moreover, we will still get the same MSE distance if we move one of the two pixels much further way. Physically similar events do not necessarily result in small MSE distances.

A completely different way to assign distance between two events is to compute the minimum ``effort'' needed to transform one image into the other, known as the optimal transport distance. 
There are many possible optimal transport algorithms (see ref.~\cite{Villani:2009} for a broad review).
Finding the minimum effort is an optimization problem:
given a cost function $c_{ij}$, where $i$ and $j$ label elements (e.g. pixel labels) of the two events, we optimize over the transport plan, $f_{ij}$.
The cost can be thought of as how much work it takes to transport a single unit of intensity a given distance, and the plan describes how much intensity to transport and where to transport it to. 
In terms of the cost and plan, the total optimal transport cost $d_\text{OT}$ is then defined as
\begin{equation}
    d_\text{OT} = \underset{f}{\min} \: \sum_{i,j} f_{ij}\, c_{ij}\:.
    \label{eq:d-ot-general}
\end{equation}
In some cases, the cost function $c_{ij}$ is itself a positive definite distance, in which case $d_\text{OT}$ is also a distance.
One example is the set of $p$-Wasserstein distances:
\begin{equation}
    d^{(p)}_\text{Wass} = \bigg(\underset{f}{\min} \: \sum_{i,j} f_{ij}\, \left(c_{ij}\right)^p\bigg)^{1/p}\:,
    \label{eq:dwass}
\end{equation}
Depending on the problem, the set of $f_{ij}$ may have to satisfy additional constraints.

We define the underlying cost $c_{ij}$ as the Euclidean distance in the $(\eta,\phi)$ plane between pixel $i$ in image $\mathcal{I}_1$ and pixel $j$ in image $\mathcal{I}_{2}$.
The transport plan $f_{ij}$ is defined by the amount of $p_T$ that is moved from pixel $i$ in image $\mathcal{I}_1$ to pixel $j$ in image $\mathcal{I}_2$.
The transport plan is constrained such that the amount of $p_T$ moved from a pixel cannot be more than what was there, $\sum_j\,f_{ij} \leq p_{T, i}$. Similarly the amount of $p_T$ moved into a pixel cannot exceed the amount in that pixel in $\mathcal{I}_2$: $\sum_i\,f_{ij} \leq p_{T, j}^{\prime}$.
Here, we consider normalized images, preprocessed such that the total intensity summed over all pixels is equal to unity, so that there is no extra cost of creating or destroying $p_T$. In mathematical language, we are considering ``balanced optimal transport''.

In particle physics applications, unbalanced optimal transport with the choice $p=1$ is commonly referred to as the Energy Movers Distance (EMD)~\cite{Komiske:2019fks, Komiske:2020qhg}, as it has the interpretation of work required to rearrange an energy pattern.
This interpretation makes the EMD a natural choice for a metric on collider events. This has prompted further work on using the EMD to define event shape observables characterizing the event isotropy~\cite{Cesarotti:2020hwb}, which can be useful in searching for signals that are far from QCD-like~\cite{Cesarotti:2020uod, Cesarotti:2020ngq}.
Sometimes, $p > 1$ has been considered~\cite{Komiske:2020qhg}, but the less explored case of $0 < p < 1$ can also be computed.
Intuitively, $p<1$ gives more importance to smaller distances. While the EMD includes an additional term to account for energy differences between jets, in our results, we will restrict to balanced optimal transport, since we normalize the images.

\begin{figure}[t]
    \centering
    \includegraphics[width=0.45\linewidth]{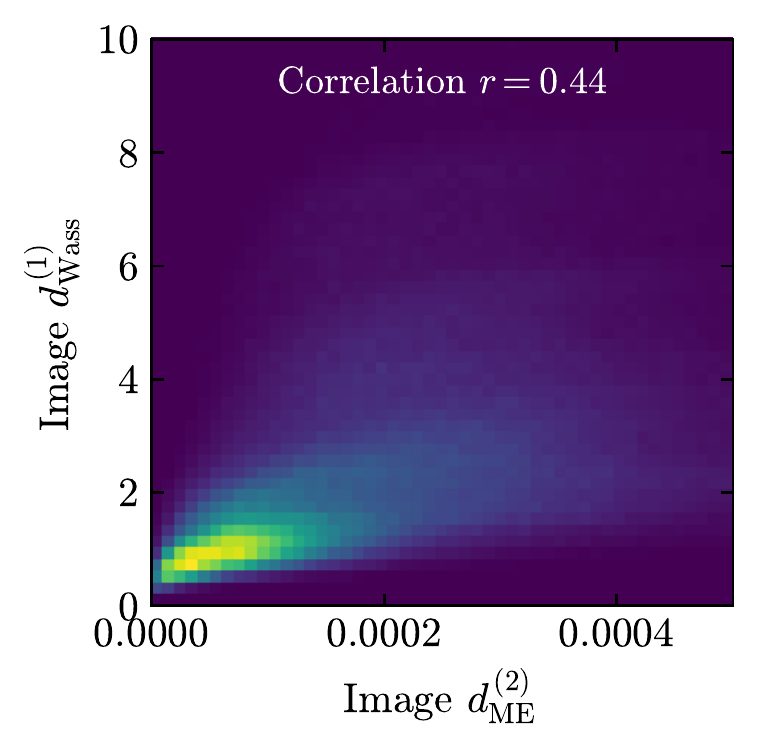}
    \caption{The pairwise (event-to-event) distances between event images for different metrics. The mean squared error $(d_{\text{MPE}}^{(2)})$ is displayed along the $x$-axis and the 1-Wasserstein distance $(d_{\text{Wass}}^{(1)})$ is along the $y$-axis.
    }
    \label{fig:Basecorrelations}
\end{figure}

The $p$-Wasserstein optimal transport metrics are more aligned with what one expects for physical events than MPE. For example, two single-particle events where the particles are nearby will have a much smaller $p$-Wasserstein distance than when they are far from each other, in contrast to their MSE distance. 
However, as QCD prefers small angle radiation, we find that the 1-Wasserstein distance and MSE have mild correlations, as shown in figure~\ref{fig:Basecorrelations}.

We reiterate that both $d_{\rm{MPE}}^{(\alpha)}$ and $d^{(p)}_\text{Wass}$ are used to compare the distance between two images (or events).
However, for anomaly detection, we want to know how far an event is from the expected distribution. One way to do this is with an autoencoder, which we describe next.

%%%%%%%%%%%%%%%%%%%%%%%%%%%%%%%%%%%%%%%%%%%%%%%%%%%%%%%%
\subsection{Autoencoders}
\label{sec:network}
%%%%%%%%%%%%%%%%%%%%%%%%%%%%%%%%%%%%%%%%%%%%%%%%%%%%%%%%

A popular method for detecting anomalous data is with a neural-network autoencoder~(AE).
An autoencoder works by first encoding the data in a lower-dimensional \emph{latent space}, and then decoding it back to the original higher-dimensional representation. The idea is that data similar to the training sample will be reconstructed well, whereas data that is not similar to the training sample may be reconstructed poorly. The reconstruction fidelity can then be used as an anomaly score. 
Often the data are represented as images, and the autoencoder uses the MSE metric (eq.~\eqref{eq:MSE-generalization} with $\alpha = 2$) to compare the input image to the reconstructed image.

\begin{figure}[t]
    \centering
    \includegraphics[width=\linewidth]{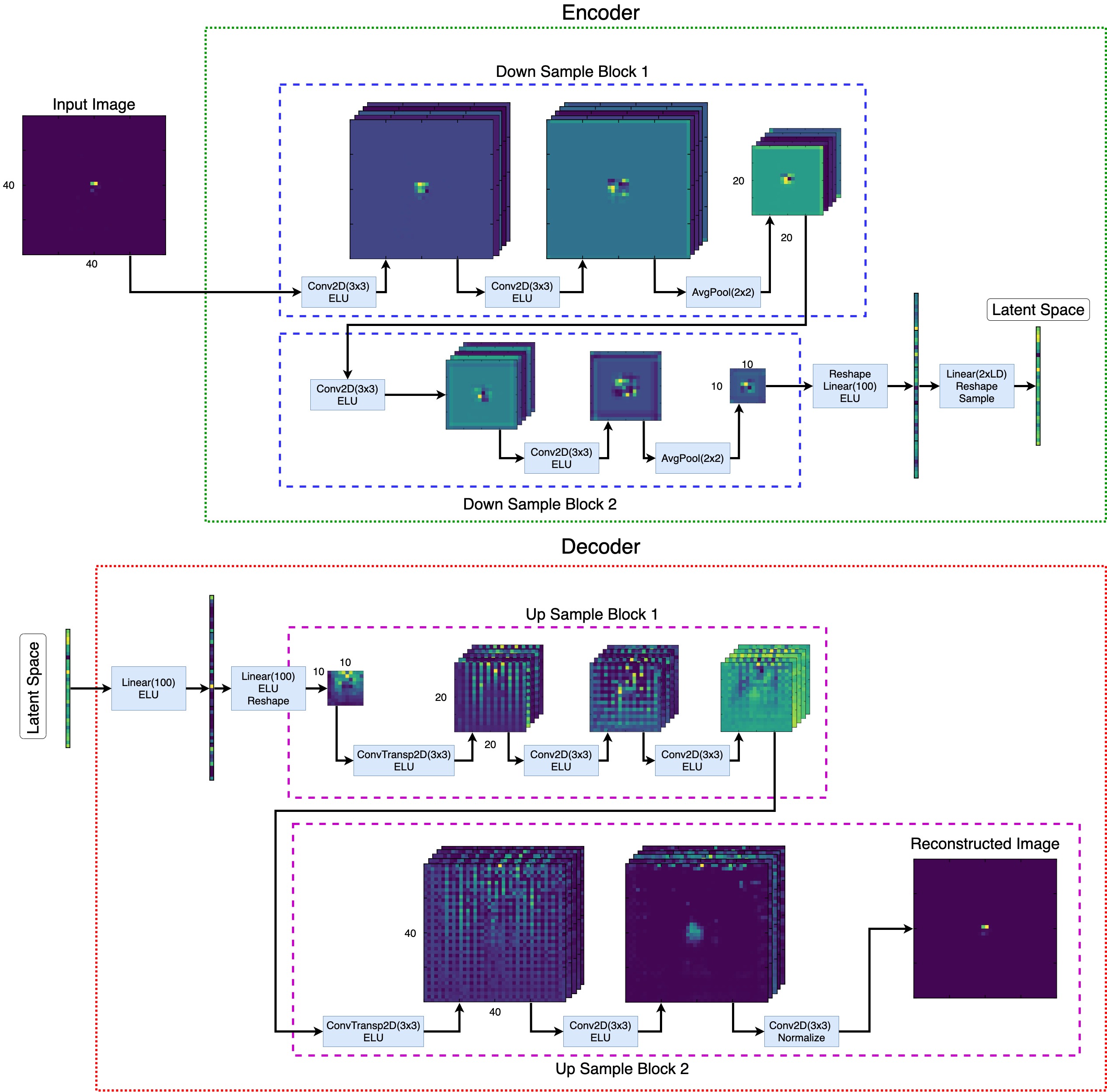}
    \caption{Example architecture of an autoencoder, as used for this study.
    The autoencoder is made of two networks, the encoder and the decoder, each with one, two, or three down(up)-sampling blocks.
    }
    \label{fig:networkArch}
\end{figure}

In figure~\ref{fig:networkArch}, we show an example of an autoencoder architecture that we will use.
The encoder is made up of some number of \emph{downsampling blocks} (there are two in the figure, each marked by a dashed blue line).
Each block contains two sets of $3\times3$ convolutional layers with a depth of five filters.
The stride and padding are set to keep the image size the same and the ELU activation function is applied after each layer.
After the convolutional layers, the data is downsampled through a $2\times2$ average pooling layer.
After the final downsampling block, the data is flattened and then followed by a dense layer with 100 nodes and an ELU activation.
Finally the network is mapped to the latent space through another dense layer.
We experiment with one, two, and three downsampling blocks, and use a fixed latent size of 64 dimensions.
Our latent space is substantially larger than what is often used, for example ref.~\cite{Farina:2018fyg} uses a six dimensional latent representation and ref.~\cite{Heimel:2018mkt} finds the optimal size to be around 20-34 for their top-tagging data.
We found the best performance with a 64 dimensional representation.

The second part of the AE is the decoder, which maps the latent space back to the space of the input data.
In our setup, the decoder is a mirror of the encoder.
The first step is a dense layer with 100 nodes and ELU activation.
From here, another dense layer is used,  where the number of nodes is set to the number of pixels in the final down sampling block.
The ELU activation function is used again, and then the data is reshaped into a square array.
From here, the same number of \emph{upsampling blocks} is applied as the number of downsampling blocks.
In each upsampling block, the first operation is a 2D transposed convolution which doubles the shape of the image and contains a depth of five filters, followed by the ELU activation.
After this, two $3\times3$ convolutional layers are used with the ELU activation with the stride and padding set to keep the image size the same.
The final convolution operation reduces the depth to one.

During training and inference, the input image is compared with the reconstructed image via some choice of event-to-event metric.
A common method is to use the MSE as the loss function, with the aim of reproducing the exact image.
However, it is possible to use other metrics for the comparison.
Furthermore, the metric used for training does not need to be the same as the metric used for the anomaly score (see for instance ref.~\cite{Cheng:2020dal}).
We will refer to the difference between the input and reconstructed image as the \emph{image distance}, also known as the reconstruction error.

A variational autoencoder enhances the basic autoencoder by adding stochasticity to the latent embedding.
In a regular autoencoder, which is a deterministic function, very dissimilar events can be placed near each other in the latent space.
Distances in the latent space of an ordinary AE therefore do not have a precise meaning.
In a VAE, the stochastic element makes the network return a distribution in the latent space for each input event. Since the same input data can be mapped to several nearby points in a VAE, dissimilar events cannot be placed nearby. 
Returning a distribution in the latent space is therefore essential for making distances in the latent space meaningful.
The stochasticity also connects the loss to the statistics method of variational inference~\cite{2013arXiv1312.6114K, 2019arXiv190602691K}, as we summarize in appendix~\ref{sec:vae} (see also refs.~\cite{Blei_2017, 2019arXiv190602691K} for reviews). Specifically, we show that the autoencoder estimates a \emph{lower bound} on the probability for any given event to be an element of the background sample that the network is trained on.

To implement the stochasticity of a VAE, our networks are trained using the standard reparameterization trick~\cite{2013arXiv1312.6114K, 2015arXiv150505770J}. A single element of the input data now yields a distribution, and these distributions are treated as a set of $D$ independent Gaussian distributions, where $D$ is the dimension of the latent space. The output of the encoder is then doubled: instead of returning a single point in the latent space, it now outputs both the means $\mu$ and the variances $\sigma^2$ of the distribution in latent space.
The loss function for the network also has to be modified: we want the background sample to be well modeled by a set of Gaussian distributions in latent space.
This is done by introducing a Kullback-Leibler divergence (KLD) term (see appendix~\ref{sec:vae} for details), which is estimated as:\footnote{
This estimation assumes Standard Normal priors for the likelihood of the latent data, as described in appendix~\ref{sec:vae}. There is a great deal of ongoing research into methods to improve the likelihood estimate by changing the latent space priors or improving the posterior approximations of the encoder~\cite{2016arXiv160604934K, 2018arXiv180305649V, Dillon:2021nxw, Aarrestad:2021oeb}.}
%boe%
\begin{equation}\label{eq:kld}
{\rm KLD} = -\frac{1}{2}\big( 1 + \log \sigma^2 - \mu^2 - \sigma^2 \big) .
\end{equation}
%eoe%
This KLD term acts to regularize the autoencoder by pushing the means in the latent space to zero and the variances to one. Depending on the metric used to determine the distance between the original and reconstructed data, more or less regularization may be needed. To account for this, we introduce another hyperparameter $\beta$, and define the loss function as
%boe%
\begin{equation}
\label{eqn:generic_loss}
L = \big(1 - \beta\big) \times \text{Image distance} + \beta \times {\rm KLD}~.
\end{equation}
%eoe%
We scan over $\beta \in \big\{ 0, 10^{-9}, 10^{-8}, 10^{-7}, 10^{-6}, 10^{-5}, 10^{-4}, 10^{-3} \big\}$, typically finding the best results for small but nonzero $\beta$.

To minimize the loss given in eq.~\eqref{eqn:generic_loss}, we use the Adam optimizer~\cite{2014arXiv1412.6980K} with the default parameters and an initial learning rate of $10^{-3}$.
The training data consists of around 550,000 QCD dijet samples, and we reserve 50,000 QCD events for an independent validation set.
After each epoch of training, the loss is evaluated on the validation set.
When the loss has not improved on the validation set for five epochs, the learning rate is decreased by a factor of 10, with a minimum learning rate of $10^{-5}$.
Training concludes when the validation loss has not improved for 12 epochs.
We then restore the weights of the network from the epoch with the best validation loss.

%%%%%%%%%%%%%%%%%%%%%%%%%%%%%%%%%%%%%%%%%%%%%%%%%%%%%%%%
\section{Autoencoder results}
\label{sec:vae-results}
%%%%%%%%%%%%%%%%%%%%%%%%%%%%%%%%%%%%%%%%%%%%%%%%%%%%%%%%

Here we present the results of our studies of variational autoencoders. We start by studying the metric dependence of VAE performance as anomaly detectors. Then we study the latent space to understand what the VAE is learning.

%%%%%%%%%%%%%%%%%%%%%%%%%%%%%%%%%%%%%%%%%%%%%%%%%%%%%%%%
\subsection{Autoencoder performance}
%%%%%%%%%%%%%%%%%%%%%%%%%%%%%%%%%%%%%%%%%%%%%%%%%%%%%%%%

Now we study the performance of variational autoencoders as anomaly detectors using different metrics. Anomaly detection with an autoencoder requires two metric choices. First, one must choose a {\bf{training metric}}, used for computing the image distance during training. Next, one must choose an {\bf{anomaly metric}} to compute an anomaly score which determines how similar an event is to the training sample. The training metric and anomaly metric can be the same, but do not have to be.

For the training metric, we consider MSE-type metrics $d_{\rm{MPE}}^{(2)}$ and $d_{\rm{MPE}}^{(1)}$ and $p$-Wasserstein metrics $d^{(1)}_\text{Wass}$ and $d^{(2)}_\text{Wass}$.
Using a $p$-Wasserstein metric in the loss function to train an autoencoder is not standard, and requires a little bit of extra engineering.\footnote{Ref.~\cite{Collins:2021pld} also implements a VAE trained with a $p$-Wasserstein metric.} The challenge is that the optimal-transport metrics are not well-suited for the back-propagation part of the training procedure of a neural network.
To get around this, we used the Sinkhorn approximation within the \textsc{GeomLoss} package~\cite{feydy2019interpolating}.
Even with this, training was slow and sometimes timed out after three days of training on GPU.
In contrast, the MSE and MAE networks typically completed training in around 12 hours on the same platform.

For the anomaly metric, we consider either using the full loss (including both the training metric contribution and the KL-divergence part in the variational autoencoder), just the MSE error between the input and output images $\big(d^{(2)}_\text{MPE}\big)$, the MAE $\big(d^{(1)}_\text{MPE}\big)$, or the $p$-Wasserstein distances $\big(d^{(\alpha)}_\text{Wass}\big)$ with $\alpha=0.5$, 1.0, and 2.0.
The value of each of these is computed for the test samples for the QCD dijet events, the top-jet events, and the $W$-jet events.

To evaluate performance in anomaly detection, we train the autoencoder on a QCD background using the training metric. Then we evaluate the anomaly score using the anomaly metric for a boosted top jet signal sample and a boosted $W$-jet signal sample. For a figure of merit of performance we use the Area Under the receiver operating characteristic Curve (AUC).

\begin{table}[htbp!]
\renewcommand{\arraystretch}{0.85}
\setlength{\tabcolsep}{6 pt}
  \centering
  \begin{tabular}{c | c c | c | c }
    \hline
    \hline
    Signal & \multicolumn{2}{c|}{} & Top jet & $W$ jet \\
    \hline
    Training  & Down & Anomaly & \multirow{2}{*}{AUC} & \multirow{2}{*}{AUC} \\
    Metric    & Samplings & Metric & & \\
    \hline
    Supervised & - & - & {\color{darkred}{\bf{0.94}}} & {\color{darkred}{\bf{0.96}}} \\
        \hline
         \parbox[t]{10mm}{\multirow{15}{*}{{MSE}}}  
         & \multirow{6}{*}{1} & Loss & 0.82 & 0.61 \\
         & & MSE &  0.82 & 0.60 \\
         & & MAE &  0.79 & 0.48 \\
         & & Wass(0.5) & 0.82 & 0.45 \\
         & & {Wass(1)} &  {0.83} & 0.41 \\
         & & Wass(2) & 0.81 & 0.39 \\
         \cline{2-5}
         & \multirow{6}{*}{2} & Loss & {0.83} & \color{darkblue}{\bf{0.65}} \\
         & & MSE & 0.83 & 0.65 \\
         & & MAE & 0.80 & 0.53 \\
         & & Wass(0.5) & 0.82 & 0.51 \\
         & & Wass(1) & 0.82 & 0.51 \\
         & & Wass(2) & 0.81 & 0.54 \\
         \cline{2-5}
         & \multirow{6}{*}{3} & Loss & \color{darkblue}{\bf{0.84}} & {0.65} \\
         & & MSE & 0.84 & 0.65 \\
         & & MAE & 0.81 & 0.53 \\
         & & Wass(0.5) & 0.83  & 0.52 \\
         & & Wass(1) & 0.84 & 0.52 \\
         & & Wass(2) & 0.82 & 0.54 \\
         \hline
         \parbox[t]{10mm}{\multirow{15}{*}{Wass(1)}}
         & \multirow{6}{*}{1} & Loss & 0.78 & 0.44 \\
         & & MSE & 0.71 & {0.57} \\
         & & MAE & 0.72 & 0.49 \\
         & & Wass(0.5) & 0.75 & 0.47 \\
         & & {Wass(1)} & {0.78} & 0.44 \\
         & & Wass(2) & 0.76 & 0.39 \\
         \cline{2-5}
         & \multirow{6}{*}{2} & {Loss} & {0.79} & 0.46 \\
         & & MSE & 0.76 & {0.61} \\
         & & MAE & 0.75 & 0.52 \\
         & & Wass(0.5) & 0.77 & 0.49 \\
         & & Wass(1) & 0.79 & 0.46 \\
         & & Wass(2) & 0.77 & 0.40 \\
         \cline{2-5}
         & \multirow{6}{*}{3} & Loss & 0.79 & 0.41 \\
         & & {MSE} & {0.79} & {0.60} \\
         & & MAE & 0.77 & 0.51 \\
         & & Wass(0.5) & 0.79 & 0.47 \\
         & & Wass(1) & 0.79 & 0.41 \\
         & & Wass(2) & 0.72 & 0.36 \\
         \hline
         \hline
    \end{tabular}
    \caption{Results showing the ability of a VAE trained on QCD only samples to distinguish top and $W$ jets as different from QCD.
    The Training Metric column shows which distance metric is used in the loss function for training, and the Anomaly Metric column shows the distance metric used at inference time.
    The bold blue entries mark the highest AUCs overall.
    We indicate the $p$-Wasserstein distance metric as Wass($p$), and the MPE with power $\alpha = 1, 2$ by MAE and MSE, respectively.}
    \label{tab:VAE_results}
\end{table}

Results are shown in table~\ref{tab:VAE_results} for the training metric choices $d_{\rm{MPE}}^{(2)}$  and $d^{(1)}_\text{Wass}$ and for different numbers of downsampling blocks in the network.
For each number of down samplings, we trained the network with different values of the VAE parameter $\beta$, and in the table present the results for the value of $\beta$ which achieved the best loss on the validation data.
For the $d_{\rm{MPE}}^{(2)}$  trained networks, the values of $\beta$ which minimized the loss were $10^{-7}$, $10^{-7}$, and $10^{-8}$, for the one, two, and three down sample block networks, respectively.
The $d^{(1)}_\text{Wass}$ trained results are in the lower part of the table and had optimal values of $\beta$ of $10^{-5}$, $10^{-8}$, and $10^{-7}$ for one, two, and three down sampling blocks, respectively.
The entries highlighted in blue indicate the configuration with the best AUC for top jets and $W$ jets across all of our considered architectures, training methods, and anomaly score methods. The top row in the table shows the AUC numbers (in red) from a supervised approach, for comparison (see appendix~\ref{sec:supervised} for details of the supervised algorithm).

In general, we find the networks trained with $d_{\rm{MPE}}^{(2)}$  as the training metric and using the full loss as the anomaly metric has the best performance.
The exception is when only a single down sample layer is used, in which case using $d^{(1)}_\text{Wass}$ as the anomaly metric does slightly better for the top-jet signal than using the full loss as the anomaly metric. 
When $d^{(1)}_\text{Wass}$ is used as the training metric, the best performance is with $d_{\rm{MPE}}^{(2)}$ as the anomaly metric. 

\begin{figure}[t]
    \centering
    \includegraphics[width=\linewidth]{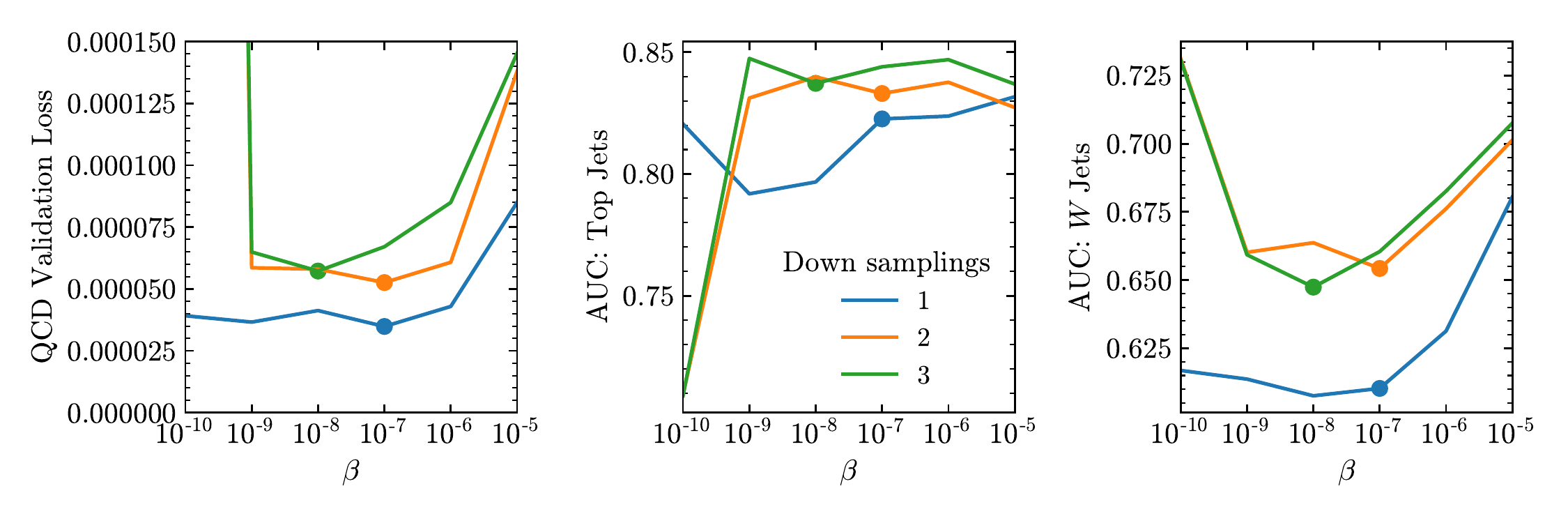}
    \caption{Results from scanning over $\beta$. 
    The value of $\beta$ which minimizes the validation loss does not yield the highest AUCs for either the top or $W$ samples.
    If one were to use one of the signal samples to chose the value of $\beta$, it can lead to worse results on the other signal.}
    \label{fig:AUC_scan_betavae}
\end{figure}

We can see at this stage the proliferation of choices one has to make when deciding what architecture, training metric, and anomaly metric to use.
Making these choices is especially hard to do if one wants to remain model agnostic.
For instance, figure~\ref{fig:AUC_scan_betavae} shows the results of the network trained with $d_{\rm{MPE}}^{(2)}$ as the reconstruction loss.
The left panel contains the loss on the QCD validation events.
Using the idea that minimizing the loss is getting a better estimate of the probability of an event, one would expect that the network configuration (number of down samplings and value of $\beta$) which minimizes the loss will have learned the QCD distribution the best.
However, the next two panels show the ability of the networks to distinguish top and $W$ jets from the QCD background.
In particular, we see that the value of $\beta$ which minimizes any of the loss curves does not yield the best signal separation.
We also point out that the network with a single down sample block has the lowest loss, but is consistently the worst anomaly detector.
This figure also highlights the danger of using the AUC of a particular signal to chose the hyperparameters of a universal anomaly detector.
Examining only performance on the $W$ jets, it would be tempting to pick either the two or three down sample networks with a value of $\beta=10^{-10}$, as this gives the best AUC for the $W$s.
However, these particular networks have the worst score of the top jets.
This is the challenge of signal independent searches; without a signal model in mind, optimizing analysis strategies is hard to do in a principled manner.

The network trained with $d^{(2)}_{\rm{MPE}}$ with a small KL divergence term yields the best anomaly detection performance. Therefore, we expect that it is learning a good representation of the underlying background distribution.
We next explore this hypothesis by examining event-to-event distances among different metrics.

%%%%%%%%%%%%%%%%%%%%%%%%%%%%%%%%%%%%%%%%%%%%%%%%%%%%%%%%
\subsection{What has the VAE learned?}
\label{sec:metrics-and-ls}
%%%%%%%%%%%%%%%%%%%%%%%%%%%%%%%%%%%%%%%%%%%%%%%%%%%%%%%%

In order for a variational autoencoder to be able to judge how likely an event is to be in a particular sample, it must have a representation of the probability distribution of events in that sample. 
Moreover, since it first maps events to a lower-dimensional latent space, the information about the relative likelihood should be encoded in the latent space in some way. 
It would make sense if the network places similar events {\it nearby} in the latent space, and dissimilar events far apart. In this section we attempt to quantify if this is indeed true by comparing to the more physical Wasserstein distance.

\begin{figure}[t]
    \centering
    \includegraphics[width=\linewidth]{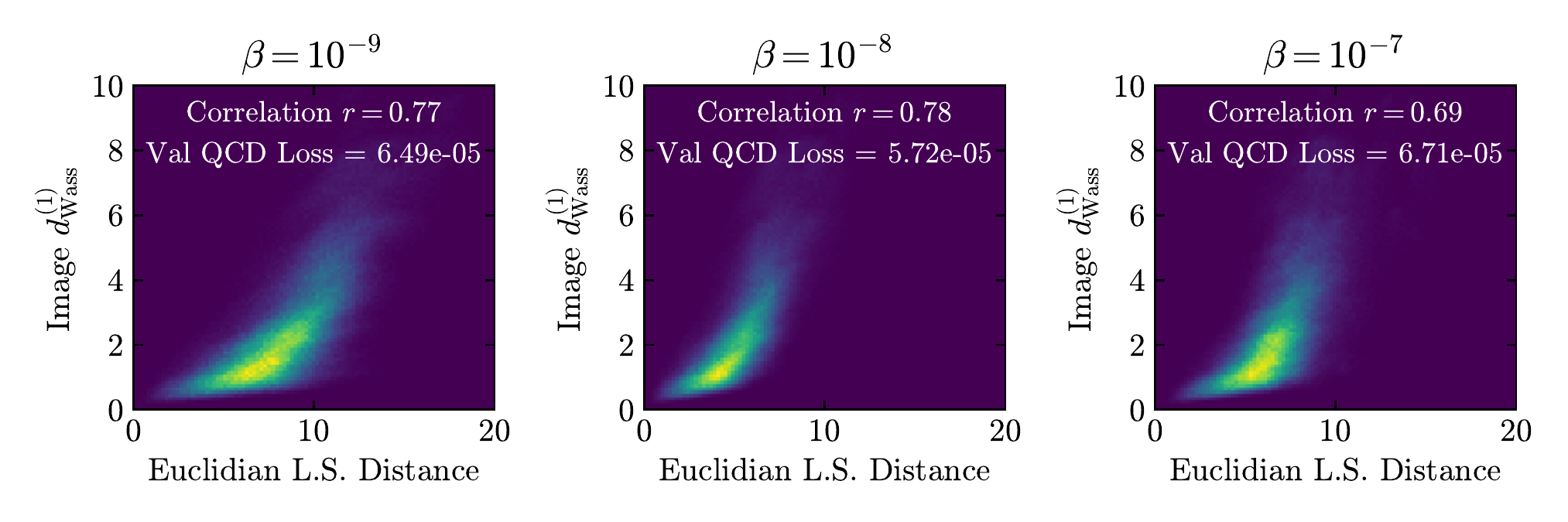}
    \caption{Each panel shows correlations between pair-wise distances of events in the QCD test set.
    The $y$-axis always denotes the \demd distance.
    The $x$-axis denotes the Euclidean distance in the latent space.
    The representation learned by the network is more correlated with the \demd distances than the MSE distances (see figure~\ref{fig:Basecorrelations}).
    }
    \label{fig:beta_scan_correlations}
\end{figure}

Since each input is mapped to a (Gaussian) distribution in the latent space, we use the Euclidean distance between the means of these distributions, which is a simple measure of the distance in latent space.\footnote{One could also try to take into account the variance of the distributions, by e.g., taking the KL divergence between the two distributions.} 
In figure~\ref{fig:beta_scan_correlations}, we
show the correlations between the Euclidean latent space distance and the $1$-Wasserstein event distance among all the $\sim 10^6$ pairings of 1000 events in the QCD test set for various values of the VAE parameter $\beta$. 
For this study, 
the events are passed through the encoder part of a VAE with three down sampling layers, down to a 64 dimensional latent space where the Euclidean distance is computed.
As the value of $\beta$ is increased, the network goes from having little regularization to being forced to approach a Gaussian.
Correspondingly, the correlation initially grows as the structure is forced upon the latent representation,
and then decreases as $\beta$ becomes so large that the regularization dominates and the distribution becomes nearly Gaussian.
We observe similar results for the networks with one and two downsampling layers that are trained with \dmse~ in the loss function.
In this figure, the value of $\beta$ which gives the minimum loss corresponds to the $\beta$ with maximum correlation, but we do not find this trend to hold in general.
It seems that the VAE with an intermediate value of $\beta$ that balances the \dmse~ and KLD terms in the loss function creates a latent space where distances between events are correlated with the \demd~ distance in the image space.

\begin{figure}[t]
    \centering
    \includegraphics[width=\linewidth]{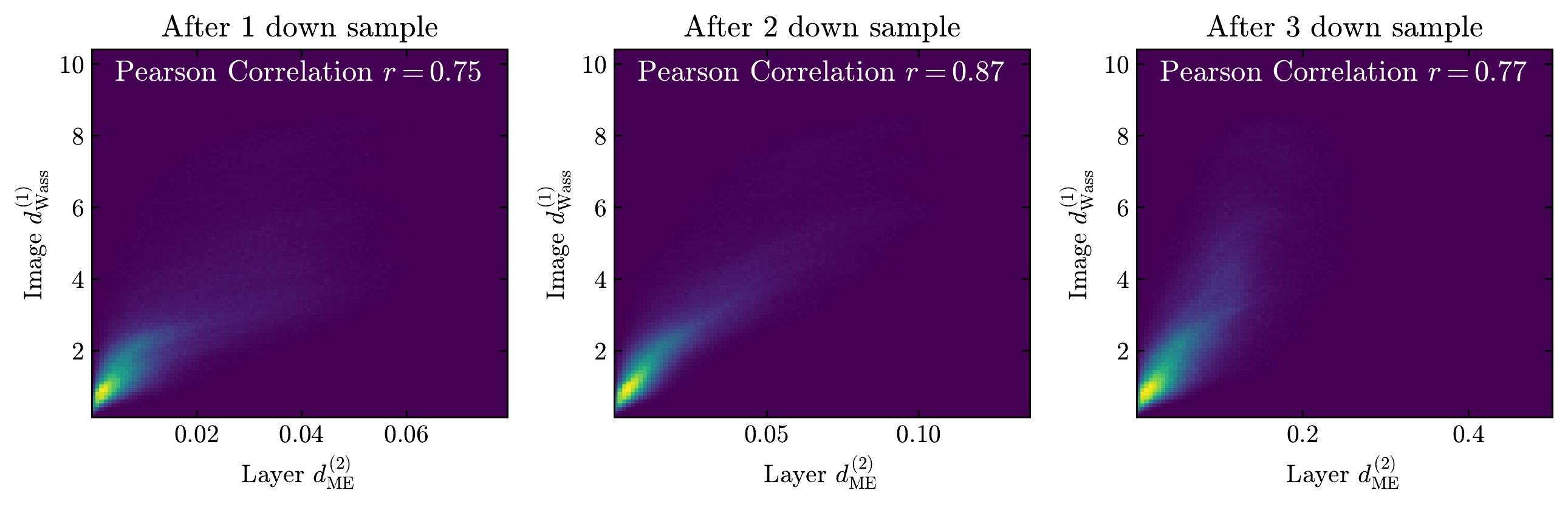}
    \caption{Correlations of the pair-wise event distance in the image space with the interior activations of the network after different number of down sample blocks.
    }
    \label{fig:downsample_correlations_p1}
\end{figure}

The downsampling operations are critical to the production of the latent space. As they combine information from neighboring pixels, they introduce an element of scale which MPE would not exhibit. 
To verify the importance of downsampling, we show in figure~\ref{fig:downsample_correlations_p1} the pair-wise event distance correlations for the same network at different depths into the encoder.
In the first panel, distances on the $x$-axis are computed in the first downsampling block, where the events are represented as $20\times20\times5$ tensors and the \dmse~ goes across all 2000 ``pixels'' (see figure~\ref{fig:networkArch}).
The correlation between the distance in this first downsampled layer and the Wasserstein distance of the events is much larger than the MSE distance between the original events. The correlation further increases from the first down sample block to the second.
The correlation then decreases after a third downsampling.
Then, when the information is further reduced to the latent space, we get smaller correlations than seen in the early stages of the network.

\begin{figure}[t]
    \centering
    \includegraphics[width=\linewidth]{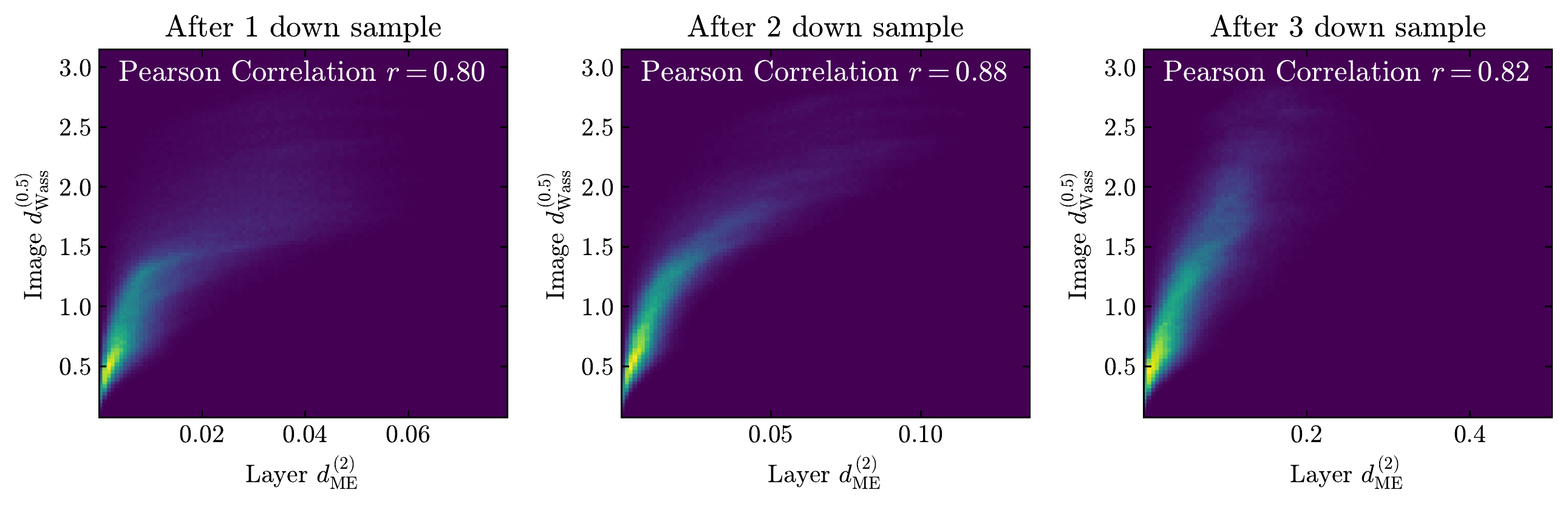}
    \caption{Correlations of the pair-wise event distance in the image space with the interior activations of the network after different numbers of down sample blocks. This is the same network as in figure~\ref{fig:downsample_correlations_p1}, but now the image distances use a different power of $p$. The correlation is higher with $p=0.5$ than $p=1$.}
    \label{fig:downsample_correlations_p_0p5}
\end{figure}

Although the EMD metric is a $p$-Wasserstein metric with $p=1$, there is no clear reason why $p=1$ should be preferred to other values. 
So, next we consider $p=0.5$.
In figure~\ref{fig:downsample_correlations_p_0p5}, we show the correlations for the same network with three down sampling layers but now using  $d^{(0.5)}_\text{Wass}$ distances along the $y$-axis.
The distances between events at different layers in the network are $\sim5\%$ more correlated with $d^{(0.5)}_\text{Wass}$ than \demd.

In this section, we have explored the representation of the QCD event distribution that variational autoencoders learn. Our conclusion from this study is that the Euclidean distances between QCD events in the latent space are highly correlated with the $p$-Wasserstein distances between the events themselves. This is particularly compelling because the VAE is trained with the MSE metric for its loss function and has no direct access to any $p$-Wasserstein metric.
A related question is how the correlations would look if a $p$-Wasserstein metric were used for training.
In that case, we find that the Euclidean distances between events in the latent space of the Wasserstein trained networks are even more correlated with the Wasserstein distances in the image space.
Thus, it could be argued that the Wasserstein trained networks learn an even better representation of the QCD distribution than the MSE trained networks.
However, the VAE with MSE training worked better for finding the top- and $W$-jet signals than those trained with an Wasserstein loss.
The fact that the method with the ``best'' latent representation does not yield the best signal separation highlights the challenges of model agnostic anomaly detection.

%%%%%%%%%%%%%%%%%%%%%%%%%%%%%%%%%%%%%%%%%%%%%%%%%%%%%%%%
\section{Event-to-Ensemble Distance}
\label{sec:metrics-based-ad}
%%%%%%%%%%%%%%%%%%%%%%%%%%%%%%%%%%%%%%%%%%%%%%%%%%%%%%%%
In the previous section we showed that VAEs tend to work better when MSE loss is used for training than when Wasserstein metrics are used for training and that the Euclidean distance in the latent space correlates strongly with the Wasserstein metric on the data, regardless of the metric used for training. Thus, there is a sense in which VAEs are learning the Wasserstein metrics. If the power of the VAE for anomaly detection in physical problems stems from it implicitly learning aspects of the $p$-Wasserstein metrics, we can then ask if there may be a way to use these metrics more directly for anomaly detection, sidestepping the VAE entirely. One way to do this is to use the metric to compute an event-to-ensemble distance, as we explore in this section.

We would like to use the $p$-Wasserstein distance, or another metric, to characterize the distance of an event to the background ensemble. There are already several options for using Wasserstein distances to characterize different types of events in the literature, such as $k$ nearest neighbor (kNN) classifiers \cite{Komiske:2019fks}, ``linearized" optimal transport~\cite{Cai:2020vzx}, where all the events are compared to a single reference event and this is used to define a new distance, and event isotropy, which compares a given event to an isotropic configuration~\cite{Cesarotti:2020hwb}. 
Our goal, using a method like these, is to extract from  the background ensemble one or more representative images and to compute the distance of a given signal or background event to those images. This direct event-to-ensemble distance measure can then be compared to the VAE anomaly score, which is also effectively an event-to-ensemble distance. 

To compute the direct event-to-ensemble distance we need an algorithm to select or construct fiducial events from the ensemble and a metric with which to compute the distance. As with the VAE architecture, there may be no choice that is optimal for all signals. In choosing the fiducial events, we must decide which sample to choose events from, how to select those events, how many events to use, how to represent the fiducial events (e.g. as images), and how to combine the distances to the different events.
To make a fair comparison to the VAE approach, we would like our algorithm for generating fiducial events to depend only on the background sample, independent of what anomalous signal we might search for. Thus we choose the QCD jet event ensemble as our reference sample. 
To select events from the sample, the simplest possibility is to arbitrarily select some number of random images. However, despite occasionally giving a large AUC for classification, results with random images are very sensitive to fluctuations between the random images. A second possibility that may seem sensible is to take the average of all events in the sample. A third option, which we find to be the most natural, is to use $k$ medoids as we now explain.

With a given metric, which we call the medoid metric, we can compute the pairwise distance $d(x_i,x_j)$ between any two events in the ensemble. Then for each event $x$ we can sum over all the distances to all other events $d(x) = \sum_{j} d(x,x_j)$.  The {\bf medoid} of the ensemble is the event $x$ that minimizes $d(x)$. {\bf $k$ medoids} generalizes this to finding the $k$ events for which the sum of the distances of each event to the closest of the $k$ medoids is minimized. Thus the event fragments into a set of clusters, with each cluster closest to one of the $k$ medoids. 
$k$-medoids clustering is similar to $k$-means clustering when the medoid metric is chosen to be the Euclidean metric, except that $k$-medoid clustering actually requires the medoid to be one of the events in the set.
Medoids have previously been explored in other contexts in refs.~\cite{Komiske:2019fks, Komiske:2020qhg, Romao:2020ojy}.
To use $k$ medoids, we need to choose a value for $k$ and a medoid metric. Then it is natural to take for the event-to-ensemble distance the distance of an event to its closest medoid. Although one could in principle use a different metric to compute the event-to-ensemble distance, it is also most natural to use the same medoid metric that determines the medoids.

\begin{figure}[t]
    \raggedright
    \includegraphics[width=0.45\linewidth]{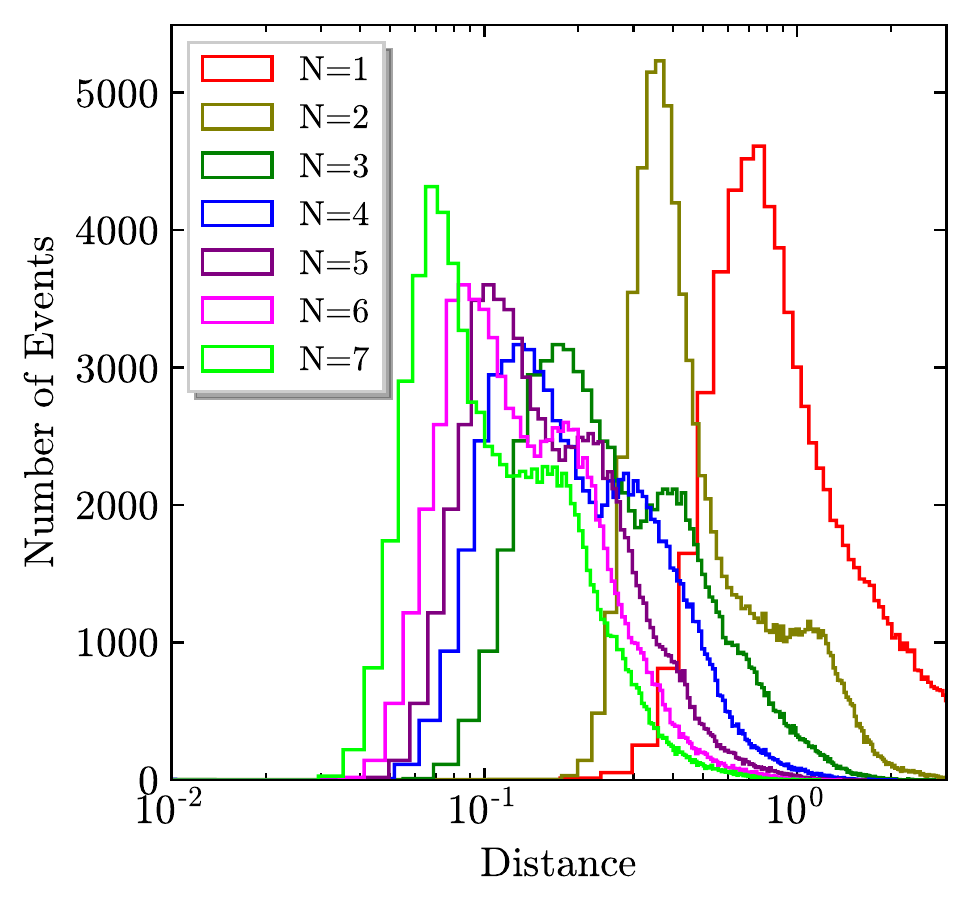}
    %\hfill
    \includegraphics[width=0.44\linewidth]{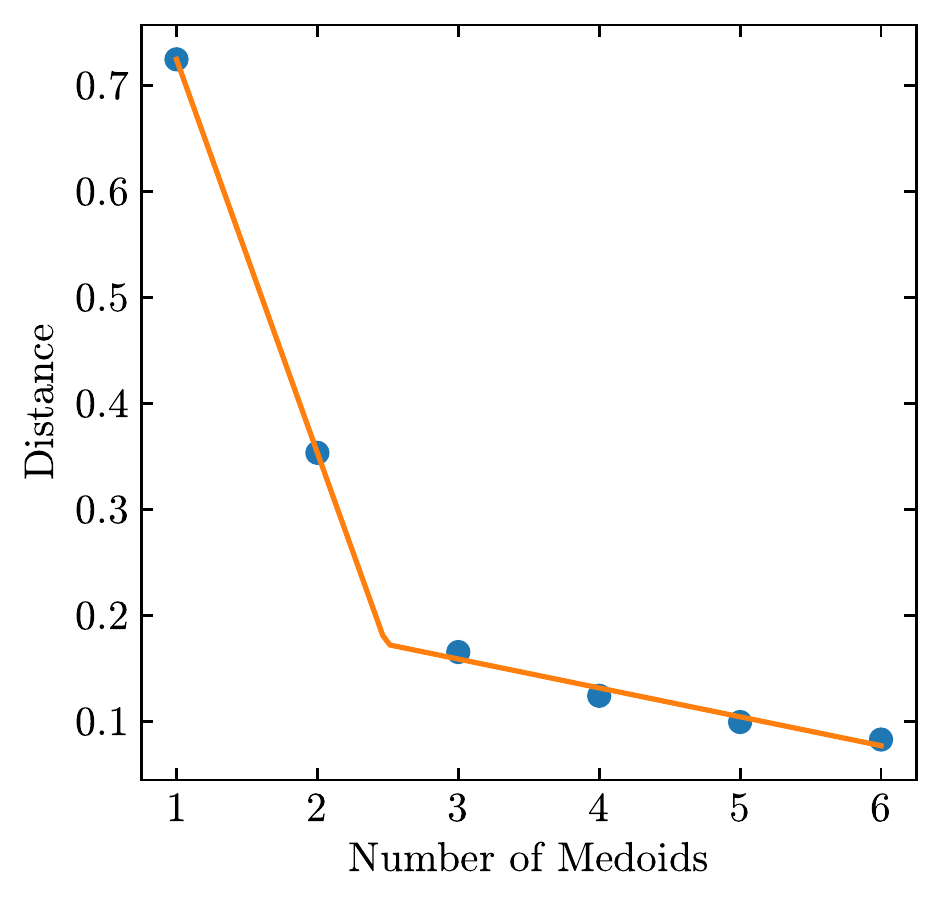}
    \caption{Example of the elbow method. Left shows histograms of the 1-Wasserstein distance to the closest medoid, with colors corresponding to the number of medoids. Right plots the peak of each of these histograms, as a function of the number of medoids.}
    \label{fig:elbow_method}
\end{figure}

Choosing the number of medoids $k$ is challenging to do in a signal independent way. 
One approach is the elbow method, a common heuristic for determining how many clusters are in a dataset. In our case, to use the elbow method we scan over the number $k$ of medoids, and for each $k$ compute the distances of all the events in our sample to the nearest medoid and histogram the results. There will be a small number events very close to a medoid and a small number very far from all medoids, so the histograms will have a peak at some finite value of the distance. Moreover, the peak distance will  decrease monotonically as the number of medoids increases. In many applications the decrease is rapid for small $k$, but at some $k$ abruptly stops decreasing rapidity and starts decreasing slowly. Thus the peak distance as a function of $k$ often has an elbow shape. To determine the elbow location algorithmically we perform  a linear regression to an elbow function (two straight lines), and take the  first integer value after the elbow as the suggested value of $k$.
The result can be seen in figure~\ref{fig:elbow_method}.
The idea behind the elbow method is that increasing $k$ past the location of the elbow should not give much improvement. Moreover, in the case of anomaly detection, if we have too many medoids, we can get one medoid that looks ``background-like" rather than ``signal-like".
We find that typically $k \sim 2-4$ medoids is selected according to this elbow method.

The main advantage of the elbow method is that it can be automated and used independent of the sample or the use case. However, there often is not a clear elbow. In figure~\ref{fig:elbow_method}, the elbow is only apparent because we have fit to a piecewise linear function. The data seems to follow more of a power law behavior. In addition, 
the location of the elbow can be affected by the maximum number of medoids we include in the fit. Additionally, the elbow can only be computed once we've already made the arbitrary choice of the medoid metric, and of the metric being used for the comparison between the full sample and the reference sample. Finally, there is no reason to expect that the elbow choice of $k$, which is determined only by the background sample, would be optimal for anomaly detection tasks. Thus, we also consider values of $k$ not determined by the elbow method for this study.

\begin{table}[h!]
\renewcommand{\arraystretch}{0.85}
\setlength{\tabcolsep}{4 pt}
    \centering
    \begin{tabular}{c| c l c | c| c }
        \hline
        \hline
        & & & & \multicolumn{1}{c|}{Top jet}  &\multicolumn{1}{c }{$W$ jet} \\
        \hline
         {\begin{tabular}{@{}c@{}}Reference \\ Sample \end{tabular}}  & Metric & {\begin{tabular}{@{}c@{}}Number of \\ medoids\end{tabular}}  & Method &  AUC & AUC\\
         \hline
Supervised & ~- & - & - & {\color{darkred}{\bf{0.94}}}  & {\color{darkred}{\bf{0.96}}} \\
        \hline
         \parbox[t]{15mm}{\multirow{15}{*}{\begin{tabular}{@{}c@{}}QCD \\ Reference\end{tabular}}}
         & \multirow{5}{*}{Wass(1)} & ~-  & Avg &  0.81 & 0.62 \\
         & & ~1  & Medoid & 0.83 & 0.66 \\
         & & ~3  (elbow)  & Medoids (min) & 0.85 & 0.68 \\
         & & ~5  & Medoids (min) &  {\color{darkblue}{\textbf{0.87}}} & 0.60 \\
         & & ~7  & Medoids (min) &  0.87 & 0.61 \\
         \cline{2-6}
         & \multirow{5}{*}{Wass(5)} & ~-  & Avg &  0.53 & 0.60 \\
         & & ~1  & Medoid &  0.68 & 0.36 \\
         & & ~3  & Medoids (min) &  0.66 & 0.41 \\
         & & ~4 (elbow) & Medoids (min) &  0.67 & 0.41 \\
         & & ~5  & Medoids (min) &  0.71 & 0.43 \\
         %& ~7  & Medoids (min) & & 0.71 & 0.46 \\
         \cline{2-6}
         & \multirow{5}{*}{MAE} & ~-  & Avg &  0.83 & 0.71 \\
         & & ~1  & Medoid &  0.82 & 0.{\color{darkblue}{\textbf{71}}} \\
         %& ~2  & Medoid & & 0.83 & 0.66 \\
         & & ~3 (elbow) & Medoids (min) &  0.82 & 0.61 \\
         & & ~5  & Medoids (min) &  0.83 & 0.67 \\
         & & ~7  & Medoids (min) &  0.83 & 0.65 \\
         \hline
         \multicolumn{6}{c}{}\\[-2mm]
         \hline 
         \parbox[t]{15mm}{\multirow{16}{*}{\begin{tabular}{@{}c@{}} Top \\ Reference\end{tabular}}}
         & \multirow{5}{*}{Wass(1)} & ~-  & Avg &  0.69 & 0.69 \\
         & & ~1  & Medoid &  0.58 & 0.79 \\
         & & ~3 (elbow) & Medoids (min) &  0.32 & 0.79  \\
         & & ~5  & Medoids (min) &  0.45 & 0.84 \\
         & & ~7  & Medoids (min) &  0.49 & 0.83 \\
         \cline{2-6}
         & \multirow{5}{*}{Wass(5)} & ~-  & Avg &  0.72 & 0.40 \\
         & & ~1  & Medoid &  0.53 & 0.52 \\
         & & ~2 (elbow) & Medoids (min) & 0.72 & 0.70 \\
         & & ~3  & Medoids (min) &  0.66 & 0.61 \\
         & & ~5  & Medoids (min) &  0.61 & 0.54 \\
         %& ~7  & Medoids (min) & & 0.63 & 0.53 \\
         \cline{2-6}
         & \multirow{3}{*}{Wass(5)} & ~3 (elbow) & Medoids (sum) &  0.66 & 0.66 \\
         & & ~5  & Medoids (sum) &  0.73 & 0.58 \\
         & & ~7  & Medoids (sum) &  0.75 & 0.60 \\
         \cline{2-6}
         & \multirow{4}{*}{MAE} & ~-  & Avg &  0.48 & 0.57 \\
         & & ~1  & Medoid &  0.29 & 0.64 \\
         & & ~3 (elbow) & Medoids (min) & 0.25 & 0.36 \\
         & & ~5  & Medoids (min) & 0.32 & 0.58 \\
         %& ~7  & Medoids (min) & & 0.33 & 0.59 \\
         \hline
         \hline
    \end{tabular}
    \caption{AUC values for QCD vs. signal classification. Top rows use a QCD reference sample, and bottom rows use a top reference sample (assuming $W$ events are more ``top-like" than QCD events). When there are multiple medoids, distances are combined either by taking the minimum or the sum to the $k$ different medoids, as denoted in the table. Medoids are selected with the same metric as is used to compare images. For each metric, we note which number of medoids corresponds to the elbow. The best AUC values with the QCD reference events are denoted in blue. 
    We indicate the $p$-Wasserstein distance metric as Wass($p$). \\} 
    \label{tab:Meds_results}
\end{table}

The top of table~\ref{tab:Meds_results} shows results for top-jet vs. QCD-jet discrimination and $W$-jet vs. QCD-jet discrimination when QCD jets are used for the reference sample. We show results for different values of $k$ with medoids, using different medoid metrics. We also show the result from using the distance to a single composite average event determined by averaging each pixel intensity over all events in the reference sample. When we study the elbow for the most common 1-Wasserstein metric, we see reasonable performance for both QCD and top jets, though it is best for neither of them. This is in line with what we expect for unsupervised anomaly detection.
For simplicity, we report results where the metric used to select the medoids is the same one used to compute our observable. We could have chosen two different metrics for the medoid metric and that used to compute the event-to-ensemble distance, but restricting to the case where they are the same does not qualitatively change our results. Table~\ref{tab:Meds_results} shows that the number of medoids and the choice of metric matters substantially.

\begin{figure}[t]
    \raggedright
    \includegraphics[width=1.0\linewidth]{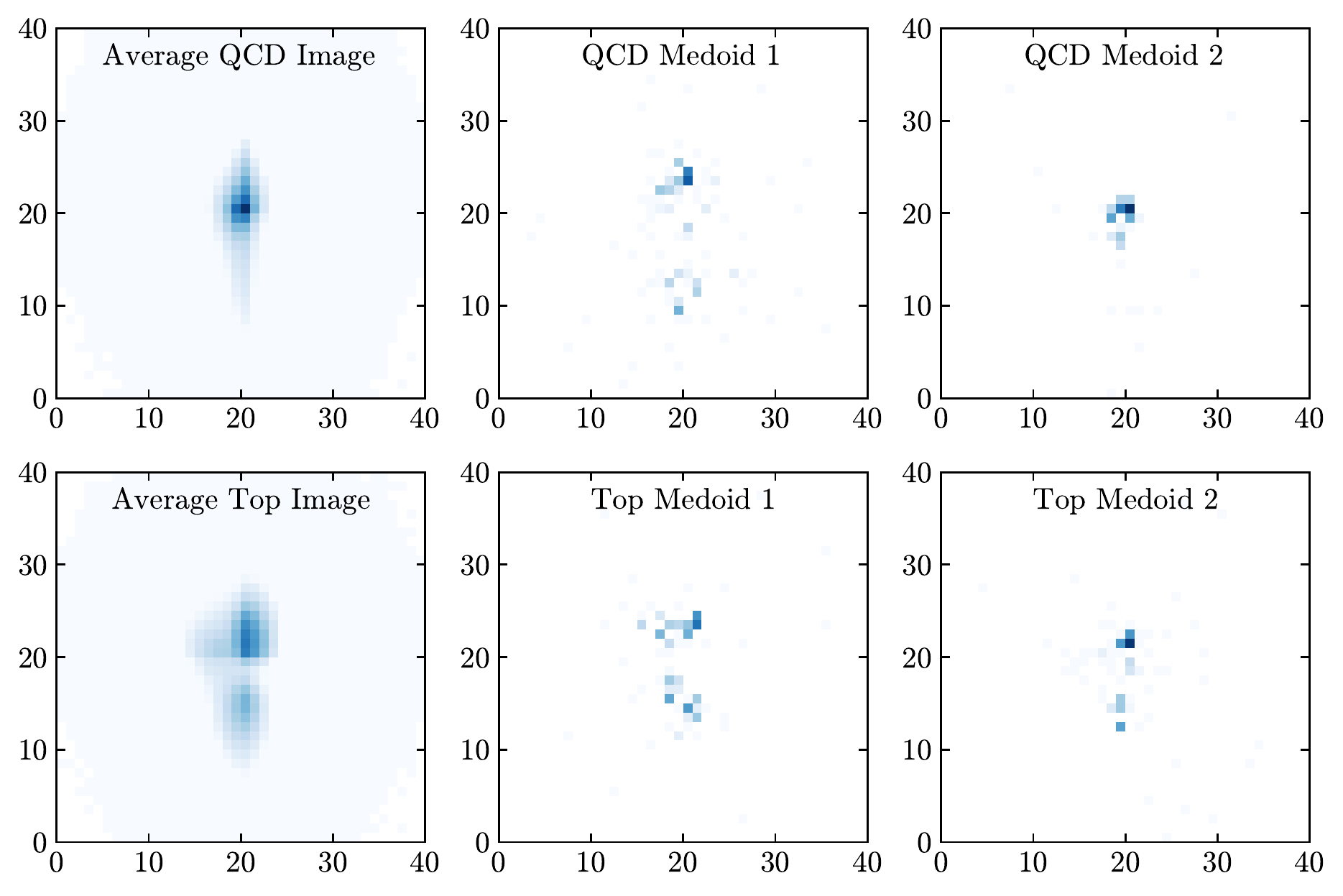}
    \caption{Images of example QCD and top events. The top row shows QCD events; the bottom row shows top events. The left column shows the average image in each case, and the other two columns show two medoids computed with the 1-Wasserstein metric. Note medoids are more sparse and varied than average images, and that one of the top medoids appears ``QCD-like" when we include multiple medoids.}
    \label{fig:ref_images}
\end{figure}

We find that the QCD medoids typically perform better than the average QCD jet. This is not surprising, since the average QCD jet is much more concentrated in the center of the image than any real QCD jet, as can be seen in figure~\ref{fig:ref_images}. In this figure, the color shows the fraction of the total $p_T$ in each pixel on a logarithmic scale. We also find better performance for anomaly detection with 5-6 medoids, rather than the 2-4 medoids suggested by the elbow method. When detecting top jets with QCD reference images, we get the best results when the $p$-Wasserstein metric with $p = 1$ is used to compare images, though we also find reasonable performance for the $p$-Wasserstein metric with $p=0.5$ or $p=2$ (not shown in the table) and when the MAE metric is used. Although MAE is not a physically motivated metric, the performance in this case is not surprising because MAE between events is highly correlated with $p$-Wasserstein distances between events for QCD jets (the Pearson correlation coefficient between MAE and the 1-Wasserstein distance is 0.87). 

That our results depend on the exponent $p$ is suggestive. For the $p$-Wasserstein metric with larger $p$ we get substantially decreased performance when comparing to QCD reference jets. This suggests that the ideal value of $p$ is related to the relevant scales in the problem: a smaller value of $p$ places comparatively larger emphasis on pixels with smaller differences. This is consistent with results such as ref.~\cite{Finke:2021sdf}, which finds better AE performance when pixel intensities are remapped to emphasize dim pixels. When we choose a better (smaller) value of $p$, the results are also slightly less sensitive to exactly which QCD reference images are chosen than when a larger value is chosen. 

While autoencoders are often trained on a QCD background, several studies have explored trying to train an AE on alternative samples. One example is ref.~\cite{Finke:2021sdf}, which showed that AEs perform poorly when tagging QCD jets if trained on a top jet sample. This can be attributed to top jets being more complex than QCD jets, so that an AE trained on top jets can also reconstruct QCD jets despite them being out of distribution samples. While modifications can be made so that an AE trained on top jets can tag QCD jets~\cite{Finke:2021sdf}, requiring sample dependent optimization defeats the point of unsupervised anomaly detection.

Unlike in the case of an AE, event-to-ensemble distances can be directly applied to other reference samples, assuming an applicable metric is selected. We include in table~\ref{tab:Meds_results} results using a top-jet reference sample for concreteness and brevity, but the method can be easily applied to other reference samples.
We find that for top reference samples, the best metric is not the same as for QCD reference samples. In contrast to the case of the QCD reference sample, we find the $p$-Wasserstein metric with higher $p$ does better for QCD vs. top classification than that with $p=1$. However, this result is signal dependent, in addition to being dependent on the background sample: when trying to use the top reference sample to distinguish QCD vs. $W$-jet events, we find using the $p$-Wasserstein metric with $p=1$ is better than higher $p$. We also find that whether the average event or minimum distance to the medoids does better depends on the signal sample --- for QCD versus top classification with a top reference sample the average event does better than medoids (unlike the QCD reference sample case), but the opposite is true for QCD vs. $W$ classification. Furthermore, we find the somewhat counter intuitive result that the sum of the distance to the medoids does better than the minimum for top vs. QCD classification with top reference jets, which is surprising because only the distance to the closest event is actually used when determining the medoids. For QCD vs. top classification, using a QCD reference sample still outperforms the top reference sample, but the opposite is true when doing QCD vs. $W$ classification.

Our best results using the event-to-ensemble distance approach
are comparable to and even slightly exceed the performance of the VAE in the previous section, which can be seen from comparing table~\ref{tab:Meds_results} to table~\ref{tab:VAE_results}. 
This suggests that if we choose a smart, physically motivated metric like the $p$-Wasserstein distance then we can use the medoid method, which is much faster and simpler than the VAE, and avoids optimizing all the hyperparameters in the VAE network architecture. 
The trade-off is that we need to put effort into optimizing the metric and choice of $k$ for the medoid approach. This is not surprising, since as we saw in the previous section, distances in the VAE latent space between two separate encoded latent representations are fairly correlated with the $p$-Wasserstein distance between the original images.
This equivalence is further supported by our study of the number of downsampling layers in the previous section. Like in the case of the $p$-Wasserstein metrics, locality is incorporated in the convolutional VAE on scales other than the arbitrary pixel size due to the convolutions and down samplings.

 The ease and speed of using the event-to-ensemble approach is a distinct advantage when compared to AEs, where the architecture, normalization, and parameters all need to be optimized. However, since the ideal metric and fiducial sample selection depend on both the background and signal samples, the signal dependence of the event-to-ensemble approach further suggests that there may be advantages of weakly/semi-supervised learning as compared to unsupervised learning, and that weakly/semi-supervised methods should be explored further. The potential advantages of semi-supervised learning can be further seen from the fact that the best QCD vs. $W$ AUC values come from top reference samples, rather than QCD reference samples.

%%%%%%%%%%%%%%%%%%%%%%%%%%%%%%%%%%%%%%%%%%%%%%%%%%%%%%%%
\section{Conclusions}
\label{sec:conc}
%%%%%%%%%%%%%%%%%%%%%%%%%%%%%%%%%%%%%%%%%%%%%%%%%%%%%%%%

Using an autoencoder for anomaly detection is particularly challenging, since it must be trained well enough to reconstruct the background, but not so well that it also reconstructs the signal. There are many details about the network configuration that need to be optimized, such as the network size, metric used to compare input and output images, the definition of the anomaly score, and hyperparameters. Many of the results currently in the literature do not sufficiently emphasize these difficulties, so we have attempted in this paper to characterize and resolve them.
To be concrete, we considered detecting boosted hadronic top and $W$ jets over a QCD-jet background.
We considered using different metrics for training the VAE; we found that using the more physically-motivated optimal transport-based metrics did not outperform the simpler mean-squared-error metrics, and actually performed slightly worse.
We found that the optimal values of various hyperparameters depend on the signal that we are trying to detect and that the optimal hyperparameters for describing the QCD sample are not necessarily those that detect anomalies the best.
	
In order to understand what the autoencoder has learned, we also studied the autoencoder latent space. The latent space provides a representation of any particular event which can be used to study the background distribution. In order to characterize this latent space, we computed the distance between distinct events. One way to do this is by using the Euclidean distance between quantities in the latent space. Alternatively, if we rely on a more physical, optimal transport based metric, we can compute the distances between images directly. When we compared the two, we found that the event-to-event optimal transport based distances between the background events are highly correlated with the Euclidean distances between events in the latent space of the autoencoder. This suggests that the autoencoder is learning some aspects of optimal transport, despite being trained with only a mean-squared-error based loss function.
	
This motivated us to develop methods that use optimal transport more directly.
By choosing a representation of the QCD background distribution, such as the average QCD image or several medoids of the set of QCD jets, we can directly compute the optimal transport distance to this fiducial sample and use it as an anomaly score. 
We found that this method is at least as effective as the autoencoder, with the added benefits of being easier and faster to optimize, and generalizing more easily than the autoencoder to more complicated background distributions. We also found that the best choice of optimal transport metric depends on both the new physics signal and the qualities of the expected background distribution.

Although we have shown that the performance of variational autoencoders can be reproduced, and improved upon, by the relatively simpler medoid method, neither approach is very close to optimal for signal detection. To be quantitative, when trained on a QCD sample, the best autoencoder performance we found gave an AUC of 0.65 for $W$ detection (see Table~\ref{tab:VAE_results}). The best performance using medoids with a QCD background gave a slightly better AUC of 0.71 (see Table~\ref{tab:Meds_results}). These are both worse than the performance of a fully supervised network which gave a nearly perfect AUC of 0.96. Somewhat surprisingly, we found that when the medoids method was used on a top-jet background sample, it found $W$ jets over QCD better (AUC of 0.84) than when trained on a QCD background. This is comparable to what a supervised network trained to find tops over QCD gives when tested on $W$ vs. QCD (AUC of 0.86). 
These observations suggest that a path forward might be to use a semi-supervised approach~\cite{Dery:2017fap, Cohen:2017exh, Komiske:2018oaa, Park:2020pak}, where a network is trained with an example signal in mind, and then used for anomaly detection more broadly.

%%%%%%%%%%%%%%%%%%%%%%%%%%%%%%%%%%%%%%%%%%%%%%%%%%%%%%%
\section*{Acknowledgments}
%%%%%%%%%%%%%%%%%%%%%%%%%%%%%%%%%%%%%%%%%%%%%%%%%%%%%%%

We thank Jack Collins, Philip Harris, and Sang Eon Park for useful discussions and comments on a previous version of this manuscript. This work is supported by the National Science Foundation under Cooperative Agreement PHY-2019786 (The NSF AI Institute for Artificial Intelligence and Fundamental Interactions, http://iaifi.org/)
The computations in this paper were run on the FASRC Cannon cluster supported by the FAS Division of Science Research Computing Group at Harvard University.
BO was partially supported by the Department of Energy (DOE) Grant No. DE-SC0020223. 
KF is supported in part by NASA Grant 80NSSC20K0506 and in part by the National Science Foundation Graduate Research Fellowship Program under Grant No. DGE1745303.
SH is supported in part by the DOE Grant DE-SC0013607, and in part by the Alfred P. Sloan Foundation Grant No. G-2019-12504.
RKM is supported by the National Science Foundation under Grant No. NSF PHY-1748958 and NSF PHY-1915071.

%%%%%%%%%%%%%%%%%%%%%%%%%%%%%%%%%%%%%%%%%%%%%%%%%%%%%%%%
\clearpage
\appendix
\section{Variational Inference for Autoencoders\label{sec:vae}}
%%%%%%%%%%%%%%%%%%%%%%%%%%%%%%%%%%%%%%%%%%%%%%%%%%%%%%%%

The idea behind variational inference for anomaly detection is to estimate the true probability distribution of the background, $p(\mathbf{x})$.
Assuming we have an underlying latent space of elements $z$, we can write $p(\mathbf{x})$ as
\begin{align}
p\left(\mathbf{x}\right) &= \mathbb{E}_{p(z)}\big[p \left(\mathbf{x}|z\right) \big]
 \equiv \int p\left(\mathbf{x}|z\right)\, p(z)\, dz,
 \label{eq:px-true}
\end{align}
where $\mathbb{E}$ denotes the expectation value, 
$p\left(\mathbf{x}|z\right)$ is the probability of $\mathbf{x}$ given $z$, and $p(z)$ is the prior likelihood of the latent data.
We can take the latent space prior to be a set of independent Gaussians with zero mean and unit standard deviation, $z_i \sim \mathcal{N}(0,1)$, where $i$ runs over the dimension of the latent space. At this point $p(\mathbf{x}|z)$ is an  unknown and intractable distribution. 

To make progress, we introduce a new \textit{tractable} distribution $q_{\phi}(z|\mathbf{x})$, where $\phi$ are some parameters to be optimized over. In an autoencoder architecture, this is the encoder. We can then write eq.~\eqref{eq:px-true} in a more useful form:
\begin{align}
p(\mathbf{x}) &= \int q_{\phi}(z|\mathbf{x})\,\frac{p(\mathbf{x}|z)}{q_{\phi}(z|\mathbf{x})}\, p(z) dz 
= \mathbb{E}_{q_{\phi}(z|\mathbf{x})}\left[\frac{p(\mathbf{x}|z)p(z)}{q_\phi(z|\mathbf{x})}\right]\:.
\label{eq:px-q-introduced}
\end{align}
The log likelihood, $\text{log}\, p(\mathbf{x})$, is then given as
\begin{align}
    \text{log}\, p(\mathbf{x}) &= \text{log}\, \mathbb{E}_{q_{\phi}(z|\mathbf{x})}\left[\frac{p(\mathbf{x}|z)p(z)}{q_\phi(z|\mathbf{x})}\right]
    \\
    &\geq \mathbb{E}_{q_{\phi}(z|\mathbf{x})}\left[\text{log}\left(\frac{p(\mathbf{x}|z)p(z)}{q_\phi(z|\mathbf{x})}\right)\right]
    \\
    &= \mathbb{E}_{q_{\phi}(z|\mathbf{x})}\left[\text{log}\, p(\mathbf{x}|z) - \text{log}\left(\frac{q_\phi(z|\mathbf{x})}{p(z)}\right)\right]\:,
    \label{eq:px-q-manipulated}
\end{align}
where we have used Jensen's inequality in the second line above. Let's first consider the first term in the last line in eq.~\eqref{eq:px-q-manipulated}. It is the expectation value of $\mathbf{x}$ given $z$ when $z$ is sampled from $q_\phi(z|\mathbf{x})$ (which is a distribution in $z$ given $\mathbf{x}$). This term can be interpreted as a (negative) reconstruction error term. If we approximate $p(\mathbf{x}|z)$ by a decoder part of the architecture $p_\theta(x|z)$ (where $\theta$ is to be optimized over), $\mathbb{E}_{q_\phi(z|\mathbf{x})}(p_\theta(\mathbf{x}|z))$ is the usual (negative) reconstruction error term in the loss function for an autoencoder with decoder $p_\theta(\mathbf{x}|z)$ and encoder $q_\phi(z|\mathbf{x})$.

The second term is by definition the Kullback-Leibler divergence (KLD) between the distributions $q_\phi(z|\mathbf{x})$ and $p(z)$. Recall that $p(z)\sim \mathcal{N}(0,1)$. We take $q_\phi(z|\mathbf{x})$ to also be a Gaussian distribution, but with a unknown mean and standard deviation (to be fixed by the optimization), i.e. $q_\phi(z|x) =\mathcal{N}(\mu(x), \sigma^2(x))$. The KLD between these two distributions is then given exactly by eq.~\eqref{eq:kld}. Using the reparameterization trick~\cite{2013arXiv1312.6114K, 2015arXiv150505770J}, we can write $q_\phi(z|x)$ in terms of a standard normal: 
\begin{align}
z \sim q_\phi(z|x)\:, 
z = \mu(x) + \sigma(x)\epsilon\:, \epsilon \sim \mathcal{N}(0,1)\:.  
\end{align}
Using the reparameterization trick allows for more efficient training of the network, as the the back propagation of the gradients extends to the parameters of the distribution ($\mu$ and $\theta$) even though a random draw from the distribution is passed to the decoder.

It's now clear that the last line in eq.~\eqref{eq:px-q-manipulated} is the negative loss for a VAE architecture. By training the VAE, we are minimizing the loss. By the inequality in eq.~\eqref{eq:px-q-manipulated}, the last line is also a lower limit for the log likelihood. The optimized VAE therefore gives a \emph{maximized} lower bound to the log likelihood, the so called Evidence LOwer Bound (ELBO).
Notice that in this discussion it is imperative to use the full VAE loss in order for it to have the variational inference interpretation. 

%%%%%%%%%%%%%%%%%%%%%%%%%%%%%%%%%%%%%%%%%%%%%%%%%%%%%%%%
\section{Supervised results}
%%%%%%%%%%%%%%%%%%%%%%%%%%%%%%%%%%%%%%%%%%%%%%%%%%%%%%%%

\label{sec:supervised}
It is well known that anomaly detection is sub-optimal for looking for any particular model; if the signal is known before-hand, supervised classification will yield the best results.
We use a similar setup for our supervised classification as we did for the VAEs.
The network consists of 1, 2, or 3 convolution blocks.
Each block is made of two successive convolutional layers with 5 filters with a kernel size of 3 pixels, followed by an ELU activation function.
After the convolutions, the data is down sampled with a $2\times2$ average pool operation.
Following the convolution blocks, the data is flattened to a vector and a fully connected layer reduces the output to a single number with a sigmoid activation.

The networks are trained using \num{50000} events from the QCD sample and \num{50000} events from either the top or $W$ samples.
Similarly, \num{5000} events from each dataset are used for validation and to stop the network training when the validation loss has stopped improving.
The training minimizes the binary cross entropy.
After training, the network is applied to the test data of \num{5000} events in each class.
We find that the network with three down sample layers achieves the best AUCs, with a score of 0.94 for top tagging and 0.96 for $W$ tagging.

\bibliographystyle{utphys}
\bibliography{metrics}

\end{document}